\documentclass{ws-ijmpa}

\newcommand{\GeV}{\ensuremath{\mathrm{Ge\kern -0.12em V}}}
\newcommand{\TeV}{\ensuremath{\mathrm{Te\kern -0.12em V}}}
\def\wangle{\mathrm{sin^{2}\theta_{eff}^{lept}(M_Z)}}

\begin{document}

\markboth{Dimitri Bourilkov}
{A Fresh Look at Gauge Coupling Unification}

%
\catchline{}{}{}{}{}
%

\title{A Fresh Look at Gauge Coupling Unification
}

\author{\footnotesize Dimitri Bourilkov}

\address{Physics Department, University of Florida, P.O. Box 118440\\
Gainesville, FL 32611, USA
}

\maketitle

\vspace{-0.35cm}
\pub{Received 26 October 2004}{}

\vspace{-0.18cm}
\begin{abstract}
The apparent unification of gauge couplings around 10$^{16}$~$\GeV$ is one
of the strong arguments in favor of Supersymmetric extensions of the
Standard Model (SM). In this contribution a new analysis, using the latest
experimental data, is performed. The strong coupling $\alpha_{s}$
emerges as the key factor for evaluating the results of the fits,
as the experimental and theoretical uncertainties in its measurements
are substantially higher than for the electromagnetic and weak couplings.
The present analysis pays special attention to numerical and statistical
details. The results, combined with the current limits on the
supersymmetric particle masses, favor a value for the SUSY
scale $\stackrel{<}{\sim}$~150 $\GeV$ and for $\alpha_{s} = 0.118-0.119$.

\end{abstract}

\vspace{-0.11cm}
\section{Introduction}

By 1991 the weak coupling was measured with much higher precision than the
strong one at LEP.
In a renowned paper~\cite{Amaldi} the famous plot was produced, showing that in
contrast to the SM the Minimal Supersymmetric Standard Model (MSSM)
leads to a single unification scale of a Grand Unified Theory (GUT),
if we let the couplings run according to the MSSM theory:
$$        M_{SUSY} = 10^{3.0 \pm 1.0} \GeV \ ;\ 
          M_{GUT}  = 10^{16.0 \pm 0.3} \GeV \ ;\ 
          1/\alpha_{GUT} = 25.7 \pm 1.7 $$
where $M_{SUSY}$ is a single generic SUSY scale where the spectrum
of supersymmetric particles starts to play a role,
$M_{GUT}$ is the scale of grand unification where the electromagnetic,
weak and strong coupling come together as an unified coupling $\alpha_{GUT}$.
In 1991, at the Z mass scale $M_Z$, the
relative errors in $\alpha(M_Z)$, $\wangle$ and $\alpha_{s}(M_Z)$
were 0.24, 0.77 and 4.6 \% respectively.

From Review of Particle Properties~\cite{RPP2004} (RPP) 2004
the relative error in
$\alpha(M_Z)$, $\wangle$ and $\alpha_{s}(M_Z)$
is  0.014, 0.065 and 1.7 \%,
so we have improved in the last decade by more than an order of
magnitude except for the strong coupling (less than 3 times).
So the time is ripe for new analyses, and one is presented
here.~\footnote{Some recent analyses are~\cite{Boer,Nima}.}

A word of caution: it is amazing that we are
trying to extrapolate from 10$^2$ to 10$^{16}$ $\GeV$.
From experiments we now that even interpolation or modest extrapolations can
be non-trivial.
We measure the ``offsets'', i.e. the value of the three couplings, around the
Z peak and rely on theory to give us the ``slopes'' of the running couplings -
without errors - up to the GUT scale.
This may be an illusion e.g. extra dimensions could modify the running already
at $\TeV$ scales, so the ``unification'' point may be imaginary.

\vspace{-0.11cm}
\section{Experimental Inputs}

From Review of Particle Properties 2004:
$$ 1/\alpha(M_Z) = 127.918 \pm 0.018 \ ;\ 
  \wangle = 0.23120 \pm 0.00015 . $$
The strong coupling $\alpha_{s}(M_Z)$ is a more complicated story,
characterized by much larger statistical errors and theory uncertainties.
Different measurements and world averages typically fall in two groups which
we will call ``high'' and ``low'' values:\\
$ \alpha_{s}(M_Z) = 0.1224 \pm 0.0038\ \ from\ \Gamma_h / \Gamma_{\mu}\ at\ Z\ peak\ (RPP\ 2004)$\\
$ \alpha_{s}(M_Z) = 0.1213 \pm 0.0018\ \ Global\ Electroweak\ Fit\ (RPP\ EW\ Section)$\\
$ \alpha_{s}(M_Z) = 0.1187 \pm 0.0020\ \ RPP\ QCD\ section $\\
$ \alpha_{s}(M_Z) = 0.1183 \pm 0.0027\ \ S.Bethke\ World\ aver.\ LEP/SLC,\ Deep\ Inel.\ Scatt.$~\cite{Bethke}

\vspace{-0.11cm}
\section{Analysis Technique}

We perform a $\chi^2$ minimization with {\tt MINUIT} for three parameters:
$M_{SUSY}$, $M_{GUT}$ and $1/\alpha_{GUT}$.
A strong correlation ($>$~0.999) between $M_{GUT}$ and $1/\alpha_{GUT}$ is
observed. The minimization problem can be re-factored with two parameters,
taking $1/\alpha_{GUT}$ as the weighted average of the three couplings at any
given scale. The results for the parameter values are numerically the same,
except for the error on the GUT coupling.
Even so the correlation $M_{SUSY}\ - \ M_{GUT}$ is $>$~0.96.

An important point is how to compute the running couplings. We perform
the analysis two times:
\vspace{-0.11cm}
\begin{itemlist}
 \item 1-loop Renormalization Group (RG) running - can solve analytically;
the coefficients for the 3 couplings are independent, given by the SM or
MSSM:
$$ 1/\alpha(\nu) = 1/\alpha(\mu) - (b_i/2\pi) \cdot ln(\nu/\mu) . $$
 \item 2-loop-RG running: additional terms so the 3 couplings depend on each
other - solved numerically; the errors depend on the scale (for typical GUT
scales they grow by 4, 12 and 6 \% respectively for the three couplings).
Here a threshold correction to $\alpha_{s}(M_{GUT})$ is applied in order to
meet the GUT boundary condition.~\cite{RPP2004}
\end{itemlist}

\vspace{-0.11cm}
\section{Results and Discussion}

The results of the fits are summarized in Table 1.
The MSSM still can provide coupling unification at a GUT scale well below
the Planck scale for the full set of precise 2004 measurements.
The strong coupling is a key for interpreting the results:
the ``high'' $\alpha_{s}(M_Z)$ values require an uncomfortably low SUSY
scale $\sim$~10 $\GeV$ -  way below the experimental lower limit
$\sim$~100 $\GeV$.
The ``low'' $\alpha_{s}(M_Z)$ values favor
values for the SUSY scale just above the present limits, e.g. for
$\alpha_{s}(M_Z)$ = 0.1187 we get $M_{SUSY} < 143(508)\ \GeV$ at 
one-sided 95 \% CL for threshold corrections of -4(-3)~\% - well in the LHC
(even FNAL or LEP2 e.g. by measuring precisely $\alpha_{s}$ above the
Z peak) direct discovery range. The results for 2-loop-RG-running with
correction -4~\% give similar values for the SUSY scale as 1-loop-RG.
The GUT scale is higher: $M_{GUT} = 10^{16.7 \pm 0.1} \GeV$, pushing the
proton lifetime ($\sim$~10$^{36}$ years) well beyond the experimental
limits ($\sim$~5 x 10$^{33}$ years).
In contrast a threshold correction of 0~\% brings the SUSY scale
to $\sim$~3 $\TeV$.

\begin{table}[th]
\tbl{Fit results - all for 2-loop-RG running except the first row.}
{\begin{tabular}{@{}lclll@{}} \toprule
Inputs            & Threshold correction & $M_{SUSY}$        & $M_{GUT}$           & $1/\alpha_{GUT}$ \\
                  &    [\%]              &[$\GeV$]           &[$\GeV$]             & \\ \colrule
\multicolumn{5}{c}{Low $\alpha_{s}(M_Z) = 0.118-0.119$} \\ \colrule
RPP 2004 QCD; 1-loop-RG
                  &  $\pm$ 0             & $10^{1.5 \pm 0.6}$& $10^{16.5 \pm 0.2}$& $23.5\pm 1.0$ \\
RPP 2004 QCD      &     - 4              & $10^{1.5 \pm 0.4}$& $10^{16.7 \pm 0.1}$ & $22.3\pm 0.7$ \\
RPP 2004 QCD      &     - 3              & $10^{2.05\pm 0.4}$& $10^{16.5 \pm 0.1}$ & $23.3\pm 0.7$ \\
RPP 2004 QCD      &  $\pm$ 0             & $10^{3.5 \pm 0.3}$& $10^{16.0 \pm 0.1}$ & $25.8\pm 0.5$ \\
S. Bethke 2002-2004
                  &     - 4              & $10^{1.55\pm 0.5}$& $10^{16.55\pm 0.15}$& $22.4\pm 0.9$ \\ \colrule
\multicolumn{5}{c}{High $\alpha_{s}(M_Z) = 0.121-0.122$} \\ \colrule
RPP 2004 EW Global Fit
                  &     - 4              & $10^{1.1 \pm 0.3}$& $10^{16.8 \pm 0.1}$ & $21.6\pm 0.55$\\
$\Gamma_h / \Gamma_{\mu}\ at\ Z\ peak$
                  &     - 4              & $10^{0.95\pm 0.6}$& $10^{16.85\pm 0.2}$ & $21.3\pm 1.1$ \\ \botrule
\end{tabular}}
\end{table}

\begin{figure}
\centerline{\psfig{file=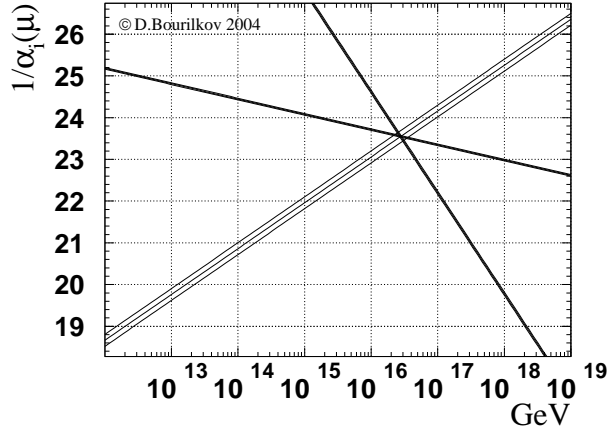,width=8.45cm}}
\vspace*{-3pt}
\caption{Example of SUSY fit with 1-loop-RG running. Perfect unification
is still possible.}
\end{figure}

\end{document}